# Domain Wall Trajectory Determined by its Fractional Topological Edge Defects

Aakash Pushp[1*§], Timothy Phung[1, 2*], Charles Rettner[1], Brian P. Hughes[1], See-Hun Yang[1], Luc Thomas[1], Stuart S.P. Parkin[1§]

[1]IBM Almaden Research Center, San Jose, California 95120, USA

[2]Department of Electrical Engineering, Stanford University, Stanford, California 94305, USA

[*]These authors contributed equally to the work.

[§]Correspondence to be addressed to: apushp@us.ibm.com (A.P.); Stuart.Parkin@us.ibm.com (S.S.P.P.)

**A domain wall (DW) in a ferromagnetic nanowire (NW) is composed of elementary topological bulk and edge defects with integer and fractional winding numbers, respectively; whose relative spatial arrangement determines the chirality of the DW. Here we show how we can understand and control the trajectory of DWs in magnetic branched networks, composed of connected NWs, by a consideration of their fractional elementary topological defects and how they interact with those innate to the network. We first develop a highly reliable mechanism for the injection of a DW of a given chirality into a NW and show that its chirality determines which branch the DW follows at a symmetric Y-shaped magnetic junction - the fundamental**

**building block of the network. Using these concepts, we unravel the origin of the one-dimensional (1D) nature of magnetization reversal of artificial spin ice systems that have been observed in the form of Dirac strings.**

The theory of topological defects has had a significant influence on the understanding of various physical phenomena ranging from superfluid Helium-3 to liquid crystals[1,2]. We study the implications of this theory in finite in-plane magnetized systems, in the shape of soft ferromagnetic NWs and their networks. Topological defects are general features in systems with broken symmetries such as head-to-head (HH) and tail-to-tail (TT) DWs[3] in these NWs. The DWs themselves have rich internal structures that can be associated with elementary topological defects[4]. Understanding the influence of the structure of the DWs on their motion in response to magnetic fields and electric currents is critical for both fundamental[5-8] and technological reasons[9,10]. Here we show first how the formation of the elementary topological defects can be manipulated to obtain DWs of a desired structure, and second how this can be used to control DW trajectory in an interconnected network of NWs. Finally, using our understanding of the elementary topological defects, we show that we can explain the formation of 1D Dirac strings commonly seen during the magnetization reversal of artificial spin ice systems[7,11].

In thin NWs with negligible intrinsic magnetic anisotropy, where the magnetization points in the plane of the NW, the competition between the exchange and magnetostatic energies leads to DWs of primarily two types[12,13] – vortex and transverse walls. In a vortex DW (Fig. 1a), the magnetization rotates by $360^o$ around a vortex core that is magnetized perpendicular to the plane of the NW, with a positive or negative polarity[14]. In a transverse DW (Fig. 1b), the magnetization rotates by $180^o$ perpendicular to the length and in the plane of the NW. Both vortex and transverse DWs can have either counter clock-wise (CCW) or clock-wise (CW) chiralities. We can use the concepts of winding numbers from the theory of topological defects to describe the magnetic texture of these different types of DWs.

The magnetic texture surrounding a topological defect can be mapped onto the order parameter space, $\theta$, defined as the angle that the magnetization makes with an arbitrarily chosen axis (Fig. 1c), as detailed in the Supplementary Information section I. For a bulk topological defect, its winding number, $n$, can then be defined[2] as:

$$n = \frac{1}{2\pi} \oint_{\partial\Omega} \nabla\theta \bullet d\mathbf{r} \qquad (1)$$

$n$ takes integer values ($\pm 1$) when $\Omega$, described by the polar coordinate $\vec{\mathbf{r}}(r,\phi)$, encompasses a bulk topological defect. For finite systems, boundary conditions play a significant role[15]. One way to incorporate the boundary conditions is by defining fractional topological defects that are confined to the boundaries of the magnetic structure under consideration, whose $n$ can be calculated[4] as:

$$n = -\frac{1}{2\pi} \int_{\partial\Omega} \nabla(\theta - \theta_\tau) \bullet d\mathbf{r} \qquad (2)$$

$n$ takes fractional ($\pm\frac{1}{2}$) values where the integration is confined to a path along the edge of the NW, and where $\theta_\tau$ is the angle of the local tangential direction along the boundary of the system.

One of the fundamental results of the topological theory of defects is conservation of the winding number of any physical system[1,2] (even when smoothly deformed). This implies that the net topological winding number of the system given by the equation below is a conserved quantity[4]:

$$n_{total} = \sum_{i}^{bulk} n_i + \sum_{j}^{edge} n_j = 1 - g \qquad (3)$$

where $g$, defined as the genus, is the number of the holes in the magnetic system. For instance, a nanodot has one bulk $n = +1$ defect in the center and no edge topological defects (Fig. 1d), whereas a nanobar has two $n = +\frac{1}{2}$ defects at the end points as shown in Fig. 1e. Likewise, in a Y-shaped junction that forms the backbone of a branched network of NWs, there are three $n = +\frac{1}{2}$ defects at the three tips (Fig. 1f). This implies that there will always be an $n = -\frac{1}{2}$ nodal defect at the junction so as to satisfy equation 3. Extending our understanding to an artificial spin ice system then leads to the realization that there are two $n = -\frac{1}{2}$ nodal defects per ring in the remnant state of a saturated honeycomb network (Fig. 1g).

One of the corollaries of equation 3 is that the winding number of a DW, $n_{total}^{DW}$, must be zero, as presence of a DW should not change $n_{total}$. For instance, a vortex DW (Fig. 1a) is composed of a bulk $n = +1$ defect in its center and two $n = -\frac{1}{2}$ defects, one at each edge of the NW, giving $n_{total}^{DW} = 0$. A transverse DW (Fig. 1b), on the other hand, is composed of an $n = +\frac{1}{2}$ defect on one edge and an $n = -\frac{1}{2}$ defect on the other edge of the NW[4,16], again with $n_{total}^{DW} = 0$. We note that the chirality of a DW is determined by the spatial arrangement of these defects. Now we consider the trajectory of DWs under the influence of a magnetic field[17,18]. When the magnetic field exceeds a critical value, namely the Walker breakdown field[19], typically between $10-20 Oe$[20,21], the DW's structure evolves as it moves. In particular, a vortex DW moves along a NW by switching its polarity back and forth (Supplementary Information section II), while preserving its chirality[18], whereas, a transverse DW moves along a NW by switching its chirality back and forth via a transient anti-vortex or a vortex DW[20]. We will show that the motion of a DW in a branched network is intimately connected to its chirality. To do this we focus on vortex DWs since their chirality is robust under motion in a magnetic field. First we need to develop a reliable method for injecting vortex DWs of a given chirality into the network.

**Controlling the chirality of injected domain walls**

We use local magnetic fields generated from current passed through an injection line to inject DWs into permalloy (Py) NWs. Although this method has been widely used[20,22,23], reliable chirality control has not yet been demonstrated. Fig. 2a shows a Scanning Electron Microscope (SEM) image of a typical device with two $500 nm$ wide injection lines separated by $6 \mu m$. The dimensions ($300 nm$ wide, $20 nm$ thick) of the Py NW, under the injection lines, are carefully chosen to favor vortex (rather than transverse) DWs[12]. We find that the placement of a notch at one edge of the NW under the injection line can be used to control the chirality of the injected DW (see Supplementary Information section III for details). In the device of Fig. 2a (inset in orange), a $60 nm$ deep triangular notch is strategically placed at the bottom edge of the NW, underneath the left injection line.

We first explore the dependence of the chirality of the injected DW on the local magnetic field strength using micromagnetic simulations. Let us consider the case when the NW's magnetization is first set in the $-x$ direction by a large magnetic field ($500 Oe$). Now the magnetic moments curl around the notch in the counter-clockwise (CCW) direction, as shown in the micromagnetic simulation in Fig. 2b[24]. It is the curvature of the magnetization that this notch engenders that favors a certain chirality of the injected DW. Applying a few nanoseconds long electrical pulse of voltage amplitude $V_{inj}$, along the injection line above a critical voltage, $V_{inj}^c$, creates a local magnetic field, $H \propto V_{inj}$, greater than the critical field, $H^c$, required to reverse the magnetization beneath the injection line. Micromagnetic simulations show that this leads to the formation of HH and TT DWs, each with CCW chirality, via intermediate states (see Fig. 2c and movie S1) composed of an anti-vortex and two vortices. It is because the cores originate near the notch at the bottom edge of the NW at $V_{inj}^c \propto H^c$ that the DWs possess CCW chirality.

We can controllably inject a DW of the opposite CW chirality by varying the strength of the injection field. As this field, i.e. $V_{inj}$, is monotonically increased, it leads to oscillatory buckling of the magnetization[25] along the NW alternating between its bottom and top edges, with a periodicity of ~2 times the NW width. Since a vortex created at the bottom edge is always CCW whereas a vortex created from the top edge is always CW (for magnetization initially along –x), this has a profound consequence that the chirality of the injected DW will oscillate with increasing $V_{inj}$. Thus, increasing $V_{inj}$ above the switching voltage $V_{inj}^{sw} > V_{inj}^c$ leads to the reliable injection of CW DWs (see Fig. 2d and movie S2; for details see

Supplementary Information sections III-IV). The results of these simulations agree well with our experimental results, as discussed below.

With a small global assisting field along the NW, $H_{inj} = 20 Oe$, applied in tandem with the local magnetic field created by a voltage pulse through the left injection line, the two injected (HH and TT) DWs move away from the injection line: one gets annihilated at the left end tip of the NW while the other gets inserted to the right of the injection line into the main section of the NW. A second $60 nm$ deep triangular 'probe notch' is positioned at the top edge in the middle of the NW to trap the injected DW (Fig. 2a inset in purple). We can experimentally determine the chirality of the trapped DW by measuring its depinning field required to dislodge the DW from the notch[26]. Fig. 2e shows the relative probability of the chirality of HH DWs injected as a function of $V_{inj}$. Below $V_{inj}^c$, no DW is found in the NW. At $V_{inj}^c = 2.6V$, HH DWs of only CCW chirality are obtained. Above the switching voltage, $V_{inj}^{sw} = 3.0V$, and for $V_{inj} < 5.0V$, the chirality of the injected HH DW is switched to CW. For $V_{inj} > 5.0V$, the chirality of the injected HH DW switches back to CCW, albeit with lower relative probability as injection becomes stochastic at high fields. TT DWs also show similar effect (Fig. 2f).

**Controlling the trajectory of injected domain walls**

We exploit the DW chirality control to demonstrate that the trajectory of a DW in a branched network is determined by its chirality. We study the DW motion in the fundamental building block of a branched network, namely a Y–shaped magnetic structure, as shown in the SEM image in Fig. 3a. The structure is formed from $20 nm$ thick Py with three branches A, B and C, each $200 nm$ wide and $6 \mu m$ long. The DW is injected from the injection line, which is positioned above a $45 nm$ deep notch (left inset Fig. 3a) in branch A. Branches B and C, angled at $\theta = 60^o$ to each other, have $60 nm$ deep probe notches that are used to trap the injected DW (right insets Fig. 3a).

We note that the initial state of the structure, even when fully magnetically saturated, has a single topological edge defect with $n = -\frac{1}{2}$ at the vertex $v_{BC}$ (see micromagnetic simulations in Fig. 3b and Supplementary Information section V). When a vortex DW is introduced, both of its fractional edge defects are constrained to move along their respective edges of the NW[4] (Supplementary Information section I), one of them leading the $n = +1$ defect and the other lagging, depending on the DW chirality. As the DW moves closer to the junction, it will travel along either branch B or C depending upon which edge has the leading $n = -\frac{1}{2}$ defect. Conservation of the total winding number implies that the vortex DW, after entering the bifurcation region, leaves its $n = -\frac{1}{2}$ defect that was behind the vortex core, on the outside vertex ($v_{AB}$ or $v_{AC}$ in Fig. 3b) of the junction, and picks up the $n = -\frac{1}{2}$ defect from the vertex $v_{BC}$ before going into the appropriate branch, determined by the $n = -\frac{1}{2}$ defect in front of the vortex core. Passage of the DW from the junction merely rearranges the location of the $n = -\frac{1}{2}$ defect from $v_{BC}$ to one of the other vertices as shown in the simulations in Fig. 3, c-f and movies S3-6. Hence, by controlling the chirality of the DW injected in branch A, the DW can be selected to enter one of the two branches B or C.

Interestingly, the DW should not only have the correct chirality but also should be of the appropriate polarity so as to pass the branched junction. This is due to the gyrotropic force[27] acting on the vortex core that pushes the $\pm$ polarity in $\mp y$ direction, for a vortex core moving in $+x$ direction. In cases where the chirality and the polarity are driven towards opposite branches (Fig. 3, d&f and movies S4&6), the polarity switches before the DW goes into the branch selected by its chirality. Thus, regardless of its polarity a CCW HH DW travels from branch A to branch C, whereas a CW HH DW travels from branch A to branch B, as summarized in the table in Fig. 3.

Fig. 4 shows the experimental verification of the chirality dependent branch selection discussed above (for a device with $\theta = 45^o$). Probability maps as a function of $V_{inj}$ and $H_{inj}$ of the DW traversing branch C ($P_C$) or branch B ($P_B$) are shown in Fig. 4 a&b respectively. As discussed earlier in Fig. 2, for $V_{inj}^c \leq V_{inj} < V_{inj}^{sw}$, a CCW chirality HH DW is injected that enters into branch C, while $V_{inj} > V_{inj}^{sw}$ injects CW chirality HH DW, which traverses branch B. Increasing $V_{inj}$ further (by $\sim 30\%$) injects predominantly CCW HH DWs, albeit with lower fidelity as the injection process becomes stochastic. In order to quantify the anti-correlation between $P_C$ and $P_B$, we define the sorting fidelity of the branched junction (Fig. 4c) as $F(\theta) = \frac{(P_C - P_B)}{(P_C + P_B)}$, whose value when $+1(-1)$ indicates that the DW traverses branch C(B) with 100% fidelity. A value of zero, on the other hand, means that the DWs travel either branch with equal probability. Events with no injection and when the DWs get trapped in the junction are discarded. Both $V_{inj}^c$ and $V_{inj}^{sw}$ decrease linearly with increasing $H_{inj}$ so that the total critical injection field, $H^c$ and the field needed for switching the chirality of the injected DW remain the same. Line cuts of the relative sorting probability into branch B and C for $H_{inj} = 50 Oe$ are shown in Fig. 4d, which mimics what is observed in Fig. 2e indicating that DW's trajectory is indeed controlled by its chirality.

The high fidelity reported here confirms that: (i) the injection process is highly chirality-controlled due to the presence of the injection notch, (ii) the vortex DWs do not switch their chirality as they traverse the junction even at fields much higher than the Walker breakdown field, (at least up to $75 Oe$ for the dimensions of our NWs) and, most importantly, (iii) the DW's trajectory through the junction is determined by its chirality, i.e., by its fractional topological edge defects.

**Origin of 1D Dirac strings in connected artificial spin-ice**

This understanding can now be extended to more complex magnetic networks, for example, to explain the propagation of magnetic monopoles in connected kagome artificial spin-ice lattices[5,7,8]. In particular, we suggest that the formation of 1D Dirac strings under the application of magnetic field in artificial kagome lattices, rather than 2D domain growth that would be anticipated from Zeeman energy minimization, is a result of the chiral nature of the DWs that propagate along one of the two possible branches depending on their constituent fractional topological defects during the switching of individual links in the honeycomb network. We illustrate the formation of two possible types of Dirac strings under magnetic field, which we define as staircase and armchair types, in an initially saturated honeycomb network. Their evolution can be understood in terms of their fractional topological defects as shown in Fig. 5 and detailed in the Supplementary Information section VI. We find that there are two distinct behaviors at the two inequivalent nodes a and b (Fig. 5a) in the hexagonal network, which depend on the location of the topological defect at the vertex of each node in the network relative to those of the DW. When the node's $n=-\tfrac{1}{2}$ defect is on the same edge as the $n=-\tfrac{1}{2}$ defect of the vortex DW (Fig. 5b) these two fractional defects annihilate together with the bulk $n=+1$ defect of the vortex DW to leave behind the $n=-\tfrac{1}{2}$ defect on the opposing edge (Fig. 5c). The second behavior is the chirality dependent branch selection that takes place at nodes a and a' whose $n=-\tfrac{1}{2}$ nodal defect is on neither edge common to those of the DW. The staircase (Fig. 5d-f) or armchair (Figs. 5d'-f') Dirac string formation then depends on transverse DW or vortex DW propagation between nodes b to a' respectively. In the transverse DW case, it is the $n=+\tfrac{1}{2}$ defect that determines its trajectory through a branched junction (Fig. 5d&e) rather than the leading $n=-\tfrac{1}{2}$ defect for the case of vortex DWs (Fig. 5d'&e' and also discussed in Fig. 3).

Using the theory of topological defects, we have obtained a detailed microscopic understanding of DW trajectories in complex branched networks. Our understanding will allow for the formation of more complex chiral magnetic orders by controllably generating and propagating several domain walls of specific chiralities into artifical spin ice structures to form defined lattices of Dirac strings. These concepts along with our capability of chirality controlled DW injection reported here can also be applied to build reliable DW based logic devices, such as a chirality controlled 2-bit demultiplexer[28], and denser racetrack memory devices that exploit the topological repulsion[23] of adjacent HH and TT DWs of appropriate chiralities.

**Figure Captions**

**Figure 1 | Characterization of topological defects. (a**, **b),** Chiral vortex and transverse HH DWs with their topological defects denoted. Polarity of the vortices is not shown. **c,** The winding number of any closed contour, $\Omega$, is calculated by measuring the total change in the order parameter, $\theta$, as the contour boundary is traversed described by the polar coordinate, $\vec{r}(r,\phi)$. **d,** A nanodot containing one bulk $n=+1$ topological defect in the center and no edge topological defects. **e,** A nanobar containing two $n=+\frac{1}{2}$ defects at the end tips. **f,** A branched Y-shaped junction with three $n=+\frac{1}{2}$ defects at each of the tips, along with one $n=-\frac{1}{2}$ defect at one of the vertices. **g,** Schematic of the remnant state of a honeycomb network magnetically saturated in the –x direction showing positions of the $n=-\frac{1}{2}$ defects in each ring.

**Figure 2 | Creation of a vortex DW of a given chirality. a,** SEM picture of a typical $300nm$ wide Py NW with the contacts separated by $6\mu m$. Left (orange) inset shows a $60nm$ injection notch underneath the left injection line at the bottom edge of the $20nm$ thick NW. Right (purple) inset shows a $60nm$ deep probe notch at the top

edge of the NW $3\mu m$ away from the left contact for DW pinning. $H_{inj}$ is applied along the x-axis. **b,** Micromagnetic simulation of a $200nm$ wide NW with a $60nm$ deep notch at the bottom edge showing the curvature of magnetization near the injection notch. **(c, d),** Time-resolved micromagnetic simulations showing the DW injection process when $51mA$ and $57mA$ currents are passed through the injection line respectively. At $51mA$ (movie S1) the vortex cores of the DWs that survive (are indicated by their polarities) originate near the notch and have CCW chirality. At $57mA$ (movie S2) the vortex cores of the DWs that survive (are indicated by their polarities) originate near the intersection of the injection line and the top edge of the NW and have CW chirality. **(e, f),** Relative injection probability for the two chiralities for HH and TT DWs respectively as a function of $V_{inj}$. For every $V_{inj}$, 100 repetitive experiments of injecting a DW and determining its chirality were performed to build statistics. Below $V_{inj}^c$ no injection takes place. Above $V_{inj}^c$ CCW HH (CW TT) DWs get injected followed by CW HH (CCW TT) DWs above $V_{inj}^{sw}$. Increasing the voltage further stochastically injects DWs of both chiralities.

**Figure 3 | Ascertaining the DW trajectory due to interplay of fractional topological defects. a,** SEM image of a branched Y-shaped junction made from $20nm$ thick Py with input branch A and output branches B and C along with their injection contact lines. Each branch is $200nm$ wide and $6\mu m$ long intersecting at the Y-junction with $\theta = 60^o$ between branches B and C. $H_{inj}$ is applied along the x-axis. Left inset shows the $45nm$ deep notch at the bottom edge of the NW underneath the

injection line A. Right insets show the probe notches in branches B and C. **b,** Micromagnetic simulation showing magnetization near the Y-junction with a $n = -\frac{1}{2}$ topological defect at the vertex $v_{BC}$ denoted by the purple dot. Vertices $v_{AB}$ (blue dot) and $v_{AC}$ (red dot) labeled as discussed in the text. **(c, d),** Time-resolved micromagnetic simulations showing CCW DWs with both polarities go into branch C with (+) polarity (see movies S3-4). **(e, f),** Time-resolved micromagnetic simulations showing CW DWs with both polarities go into branch B with (-) polarity. (see movies S5-6). The results of the simulation are summarized in the truth table. Topological defects with their respective $n$ indicated. All vortex cores have an $n = +1$ topological defect (not indicated) regardless of their polarity. In **d** and **f** the $n = -1$ defect in reality signifies two $n = -\frac{1}{2}$ defects along the edge on both sides of the vertex $v_{BC}$, which form when the $n = +1$ defect gets created in the bulk.

**Figure 4 | Experimental verification of the DW chirality based trajectory. (a, b),** Probability maps as a function of $V_{inj}$ and $H_{inj}$ for the vortex HH DWs traversing branch C ($P_C$) and branch B ($P_B$) respectively. For each $V_{inj}$ and $H_{inj}$, 50 repetitive experiments of injecting a DW and determining which branch of the magnetic structure the DW traversed were performed to build statistics. **c,** Sorting fidelity $F(\theta)$ as a function of $V_{inj}$ and $H_{inj}$ for devices with $\theta = 45^o$ and a $45nm$ deep injection notch. Red (blue) region implies DWs are going into branch C (B). **d,** Relative probability of finding HH DWs in branches B and C after having been

injected from branch A as a function of $V_{inj}$ and $H_{inj} = 55 Oe$ indicated by the dashed line cuts in **a** and **b**.

**Figure 5 | Unraveling the origin of 1D Dirac strings in artificial spin-ice.** Micromagnetic simulations depicting the growth of Dirac strings in an artificial spin ice kagome lattice under the application of a magnetic field. The topological defects of the DWs and at the vertices of the nodes of the network are denoted by $\pm\frac{1}{2}$ and +1. The two sets of inequivalent nodes are indicated in **a**, by a, a' and b, b'. Upper right insets in each panel show schematically the DW trajectory in the network. **(a-c),** We consider a CCW DW entering node a with the nodal $n = -\frac{1}{2}$ defect at the vertex away from the edges containing the DW's fractional topological defects. In this case, the chirality dependent branch selection occurs (just as in Fig. 3). The DW afterwards travels along the branch connecting nodes a and b. At node b, the leading $n = -\frac{1}{2}$ and the $n = +1$ defects of the DW merge with the nodal $n = -\frac{1}{2}$ as the two fractional defects share the same edge. **(d-f),** Evolution of a staircase Dirac string when a CW transverse DW is formed in the link between nodes b and a'. Middle inset in **f** shows schematically the staircase structure formed in the honeycomb network when the events depicted in **(a-c)** and **(d-f)** repeat. **(d'-f'),** Trajectory of an armchair Dirac string occurs when a CW vortex DW is formed in the link adjoining nodes b and a'. The middle inset in **f'** illustrates the armchair Dirac string formed when the events depicted in **(a-c)** and **(d'-f')** repeat.

**Author contributions:** A.P. and T.P. conceived, performed the experiments, and analyzed the data. T.P. did the micromagnetic simulations. S.-H.Y. grew the films. C.R. and B.P.H. did nanofabrication of the devices. A.P., T.P. and S.S.P.P. wrote the manuscript. S.S.P.P. supervised. All authors discussed the results and implications.

**Correspondence:** apushp@us.ibm.com (A.P.); Stuart.Parkin@us.ibm.com (S.S.P.P.).

**Acknowledgements:** We gratefully acknowledge discussions with O. Tchernyshyov, S. A. Parameswaran, K. P. Roche, R. Roy, S. Raghu, S. Kivelson and N. P. Aetukuri.

# Figure 1

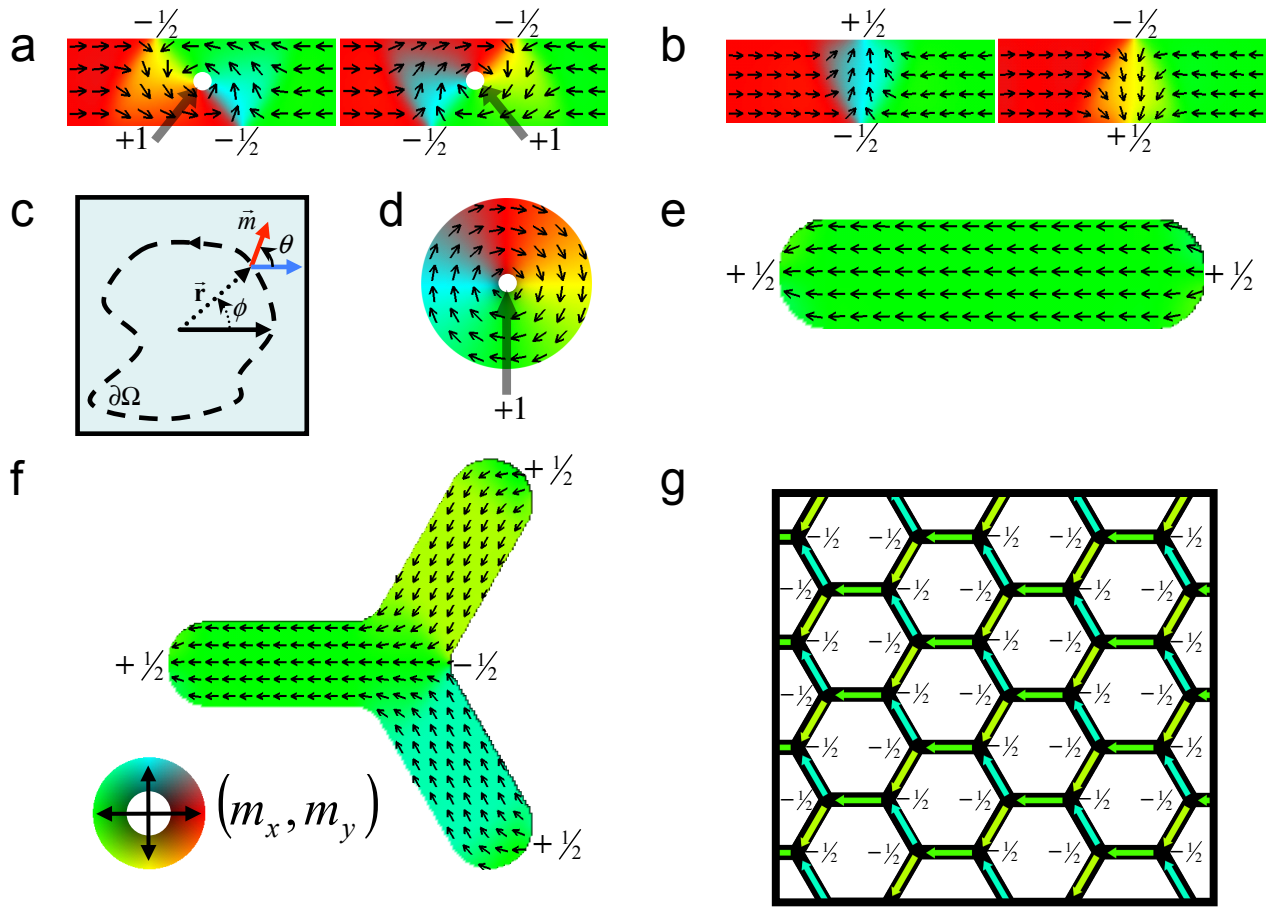

Fig. 1 Pushp, Phung et al.

# Figure 2

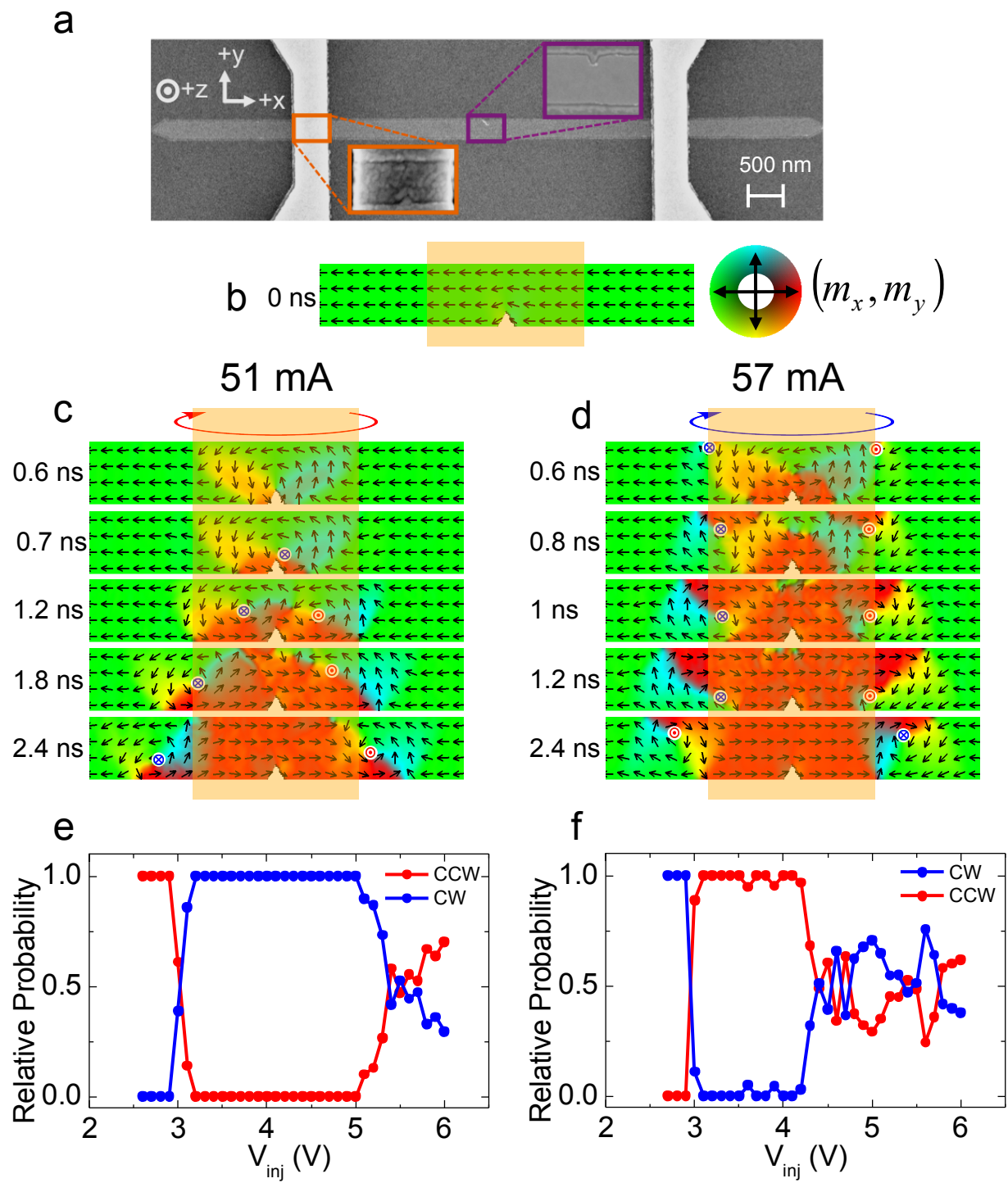

Fig. 2 Pushp, Phung et al.

**Figure 3**

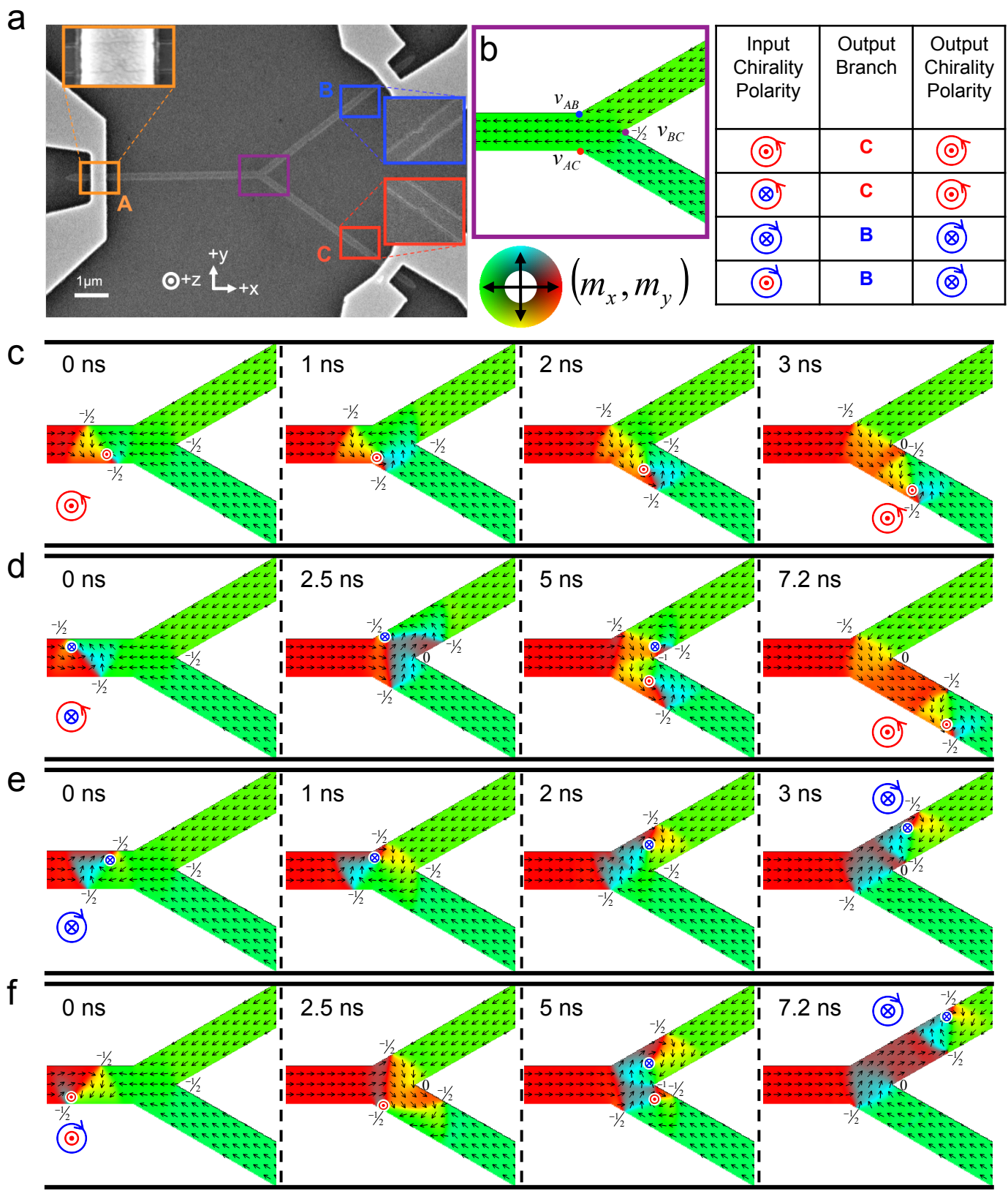

Fig. 3 Pushp, Phung et al.

# Figure 4

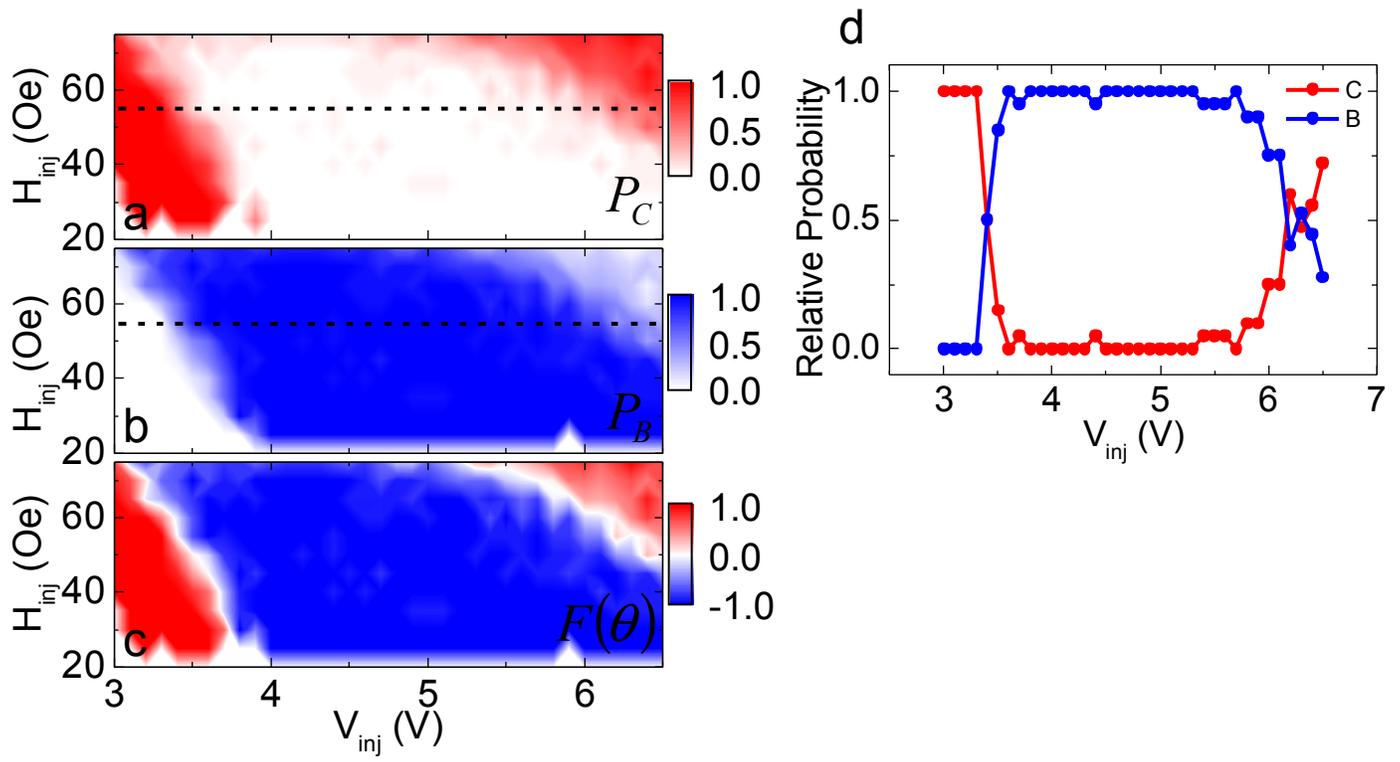

**Figure 5**

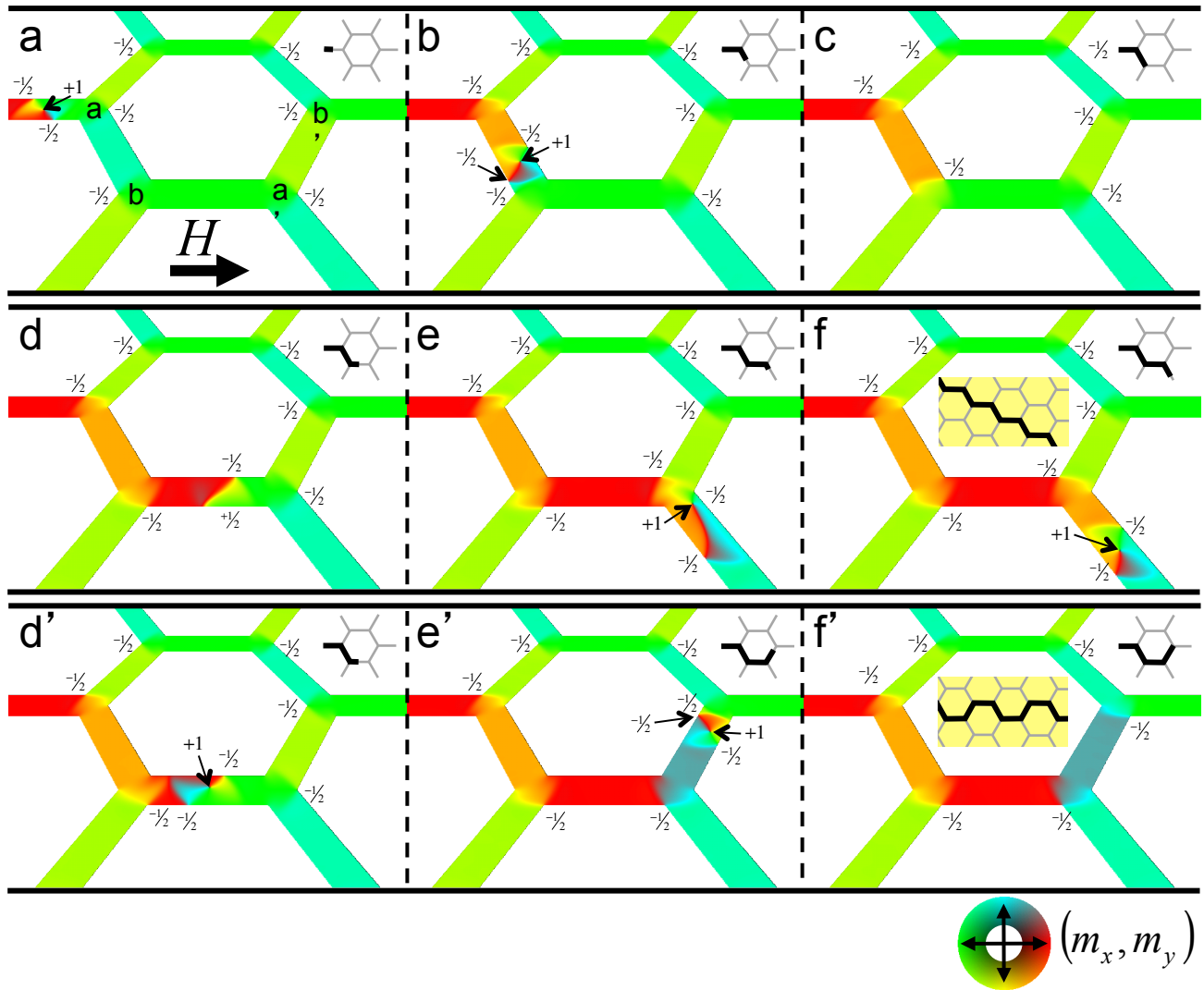

Fig. 5 Pushp, Phung et al.